# Multimodal Affect Analysis for Product Feedback Assessment

**Amol S. Patwardhan and Dr. Gerald M. Knapp**
MIE, LSU

## Abstract

Consumers often react expressively to products such as food samples, perfume, jewelry, sunglasses, and clothing accessories. This research discusses a multimodal affect recognition system developed to classify whether a consumer likes or dislikes a product tested at a counter or kiosk, by analyzing the consumer's facial expression, body posture, hand gestures, and voice after testing the product. A depth-capable camera and microphone system - Kinect for Windows – is utilized. An emotion identification engine has been developed to analyze the images and voice to determine affective state of the customer. The image is segmented using skin color and adaptive threshold. Face, body and hands are detected using the Haar cascade classifier. Canny edges are identified and the lip, body and hand contours are extracted using spatial filtering. Edge count and orientation around the mouth, cheeks, eyes, shoulders, fingers and the location of the edges are used as features. Classification is done by an emotion template mapping algorithm and training a classifier using support vector machines. The real-time performance, accuracy and feasibility for multimodal affect recognition in feedback assessment are evaluated.

## Keywords

## 1. Introduction
People rely on multiple modalities to express and identify affect (feelings and emotions). There has been considerable research on unimodal or bimodal affect recognition in the past decade. Bartlett et al. [1] provided automatic facial action detection results from spontaneous expressions. Another work [2] performed automated facial expression recognition in real time based on optical flow. Research on bimodal affect recognition [3-5] has focused on using body, hand gesture or speech as the second modality in addition to face. The recognition rates using more than one modality have been reported to be higher than unimodal affect recognition rates [6]. The study of multimodal affect recognition has gained significant attention of researchers in recent years. Experiments done in [7] aimed at detecting interest level of child solving a puzzle by using sensory information from face and body posture. Spontaneous smiles were analyzed using multiple modalities in a work done by Cohn et al. [8]. Fusion of data from face, body and speech at feature level and result level was compared [9]. Work done in [10] discussed effects of user interaction on automated affect recognition. Multimodal affect analysis was performed on speech interaction with a user agent and a Bayesian classifier was used in a study [11]. Audio visual cues and application of multimodal affect recognition for interactive interfaces was assessed [12]. Emotion synthesis using a 3D virtual agent was evaluated in [13] and facial, gesture and posture were used for emotion recognition. Despite the ongoing research, very few automated multimodal affect recognition systems exist because of the challenges such as reliable fusion techniques, noise and processing speed as discussed in a survey [14].

This research paper performs an initial study on a multimodal affect recognition system containing two main components. The first component is a frame based feature point, skeletal joint, image processing and speech recognition component and the second component is responsible for offline affect recognition of the tracked feature data. The system used multiple modalities such as video, speech, facial expressions, body posture, head and hand position as the input. The study used precision, recall, tracking processing time as the metrics for the assessment of tracking and affect recognition components using multiple modalities. Furthermore, the paper evaluated the feasibility of such a multimodal affect recognition system and its application in obtaining product feedback from customer behavior. A multimodal affect recognition system was developed to determine if the customer exhibits negative affect such as being unhappy, disgusted, frustrated, angry or positive affect such as happy, satisfied and content with the product being offered. The system can be used in four different scenarios involving product feedback. For instance, multimodal affect recognition system can be used for feedback assessment when a customer tests a sample of food, perfume, jewelry, shoes or sunglasses because these products invoke strong emotions from the person which are conveyed through facial expressions, body posture, head and hand position. Additionally, we consider a scenario where the customer returns a product after testing it at home. The body language and head position can serve as cue whether the customer is returning the product because he is unhappy with the product or simply because it did not match their needs. A third scenario is a restaurant where the customers may provide hints about whether they liked the cuisine or not, using facial expression, head nod and body posture. Finally, the multimodal affect recognition system can prove useful to detect an angry, frustrated customer at a ticket counter or while waiting in line by analyzing the body posture, facial expression, head position and hand gestures and speech. The ability to identify product feedback based on automatic multimodal recognition of affect can prove useful for determination of popular products and customer preferences. The feedback data can be used by manufacturers to decide future product lines, designs and pricing.

A study done in [15] has shown that people display 6 basic emotions (anger, disgust, sadness, happiness, fear and surprise) and facial expressions, body posture, hand and head positions provide cues for recognizing these basic emotions. Thus, we incorporated these basic emotions as indicators to determine the customer affect after testing a product. The research examines automatic disgust and anger recognition and associates the two emotions to dissatisfaction with a product. A person may exhibit disgust after tasting a bad recipe, after smelling an awful perfume or a stinking product. A customer may display anger and frustration at a return counter because of a faulty electronic device or delayed restaurant service or show annoyance towards terrible hotel room service at a checkout counter. On the other hand, the research focuses on automatic recognition of smile and associates the emotion to overall customer satisfaction. Customers often smile because they are happy with the sunglasses they tested, or after test driving the luxury car of their dreams, or while handling a newly launched notebook and a cell phone. Thus, an important objective of this research is to improve smile, disgust and anger recognition using multiple modalities due to the importance of these 3 emotions in product feedback. In this research the following issues have been investigated: First, we measure the precision, recall and rate of the recognition of affect using individual modalities such as facial expression, head and hand position, body posture and speech and then compare the results with results obtained using multimodal affect recognition. Second, geometric features and Facial Action Units (FACS) have been used in studies [16, 17] to successfully detect emotions from facial expression and other modalities such as hand and head. Thus, this research uses the skeletal and facial tracking available in depth sensing camera such as the Kinect sensor, to train a classifier using support vector machines (SVM) and measures the accuracy and speed for affect recognition using supervised learning. The goal of the research is to implement a prototypical multimodal affect recognition system that uses Kinect sensor. Third, the research proposes a key word based lookup technique for affect recognition from speech and an emotion template mapping algorithm for affect recognition from facial images. The template based algorithm determines the affect based on a set of rules as opposed to supervised learning based approach. The study uses snapshot of an image from the video and feeds it as an input to a module that uses the emotion template algorithm for affect recognition. The edge length, location, orientation and count are analyzed to determine affect. The analysis is performed on face detected by Haar cascade classifiers and segmentation based on skin color and adaptive threshold. One of the research objective is to find evidence whether template based mapping algorithm and the key word lookup technique improve or complement the accuracy results from traditional supervised learning approach. Finally, the study measures the processing times for tracking multiple modalities to determine feasibility of the real-time usage of the multimodal affect recognition system.

The rest of the paper has been organized into the following parts: Section 2 describes the features used in facial expression tracking, head position, hand position and body posture tracking. Section 3 explains the emotion template algorithm approach for affect recognition and its use in image snapshot analysis. The section also focuses on a key word lookup based approach for affect recognition from speech. Section 4 contains details about the prototype program for multimodal affect recognition system. Section 5 provides the results of experiments. Section 6 discusses the conclusions based on the research.

## 2. Features

### 2.1 Facial Expression

Facial expressions were tracked using the face tracking API available in the Microsoft Kinect Software Development Toolkit. Currently a total of 121 facial points are available for tracking through the toolkit. Out of the 121 available points only the 60 non-rigid points around eyebrows, eyes, mouth and cheeks were tracked. The feature vector was defined using the technique described in [17]. In this technique, the geometric features such as distance, angle with the horizontal axis and the co-ordinates of tracked points are used. For the feature vector from facial expression modality, the screen coordinate location of the tracked points, the Euclidean distance between each pair of points, the angle between the straight line passing through each pair of points and the horizontal axis was used. The feature vector for the facial expression modality can be defined as follows:

$$FV_f = \{P_{1,x}(n), P_{1,y}(n), \ldots P_{60,y}(n), d\ P_1(n), P_2(n), \ldots, d\ \theta\ P_1(n), P_2(n), \ldots, \theta\ P_{59}(n), P_{60}(n) \quad (1)$$

The value $P_{i,c}(n)$ is the x, y co-ordinate value of $i^{th}$ tracked point in the $n^{th}$ frame where $c \in \{x, y\}$. d is the Euclidean distance between two distinct tracked points. $\theta$ is the angle between two distinct tracked points.

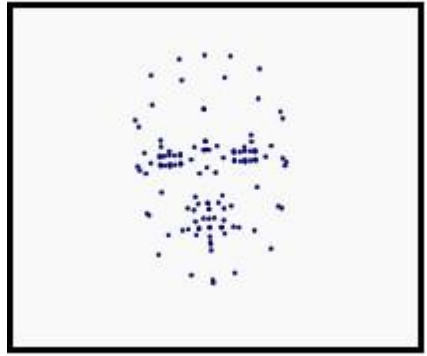

Figure 1: Facial expression tracking.

### 2.2 Head Position

The head position was tracked using the face tracking API available in the Microsoft Kinect Toolkit. For head position tracking 12 points were used described as follows: Top of skull, leftmost side of skull, rightmost side of skull, top left point on forehead, top right point on forehead, top center point on nose, point on tip of nose, point on left cheek, point on right cheek, point on left side of chin, point on right side of chin and the point on bottom part of chin. For the feature vector from head position modality, the screen coordinate location of the tracked points, Euclidean distance between each pair of points, the angle between the straight line passing through each pair of points and the horizontal axis was used. The feature vector for the head position modality can be defined as follows:

$$FV_{he} = \{P_{1,x}(n), P_{1,y}(n), \ldots P_{12,y}(n), d(P_1(n), P_2(n)), \ldots, d(P_{11}(n), P_{12}(n)), \theta(P_1(n), P_2(n)), \ldots, \theta(P_{11}(n), P_{12}(n))\} \quad (2)$$

The value $P_{i,c}(n)$ is the x, y co-ordinate value of $i^{th}$ tracked point in the $n^{th}$ frame where $c \in \{x, y\}$. d is the Euclidean distance between two distinct tracked points. $\theta$ is the angle between two distinct tracked points.

### 2.3 Hand Position

The hand position was tracked using skeletal tracking available in the Microsoft Kinect API. For hand position tracking 8 points were used. The left shoulder, left elbow, left wrist, left palm, right shoulder, right elbow, right wrist and right shoulder were tracked. For the feature vector from the hand position modality, the screen coordinate location of the tracked points, Euclidean distance between each pair of points, the angle between the straight line passing through each pair of points and the horizontal axis was used. The feature vector for the hand position modality can be defined as follows:

$$FV_{ha} = \{P_{1,x}(n), P_{1,y}(n), \ldots P_{8,y}(n), d(P_1(n), P_2(n)), \ldots, d(P_7(n), P_8(n))$$
$$, \theta\ P\ (n), P\ (n)\ , \ldots, \theta\ P\ (n), P\ (n) \quad (3)$$

The value $P_{i,c}(n)$ is the x, y co-ordinate value of i tracked point in the n frame where c ϵ {x, y}. d is the Euclidean distance between two distinct tracked points. $\theta$ is the angle between two distinct tracked points.

## 2.4 Body Posture

The body posture was tracked using skeletal tracking available in the Microsoft Kinect API. For body posture tracking, only the region above waist was considered. This is because in most cases the system will be used over the counter where only the upper body is visible and the person's lower body would be occluded. A total of 14 points, including 8 points previously discussed for hand position tracking and 6 points consisting of head, shoulder center, spine, hip center, left hip and right hip were tracked. For the feature vector from body posture modality, the screen coordinate location of the tracked points, Euclidean distance between each pair of points, the angle between the straight line passing through each pair of points and the horizontal axis was used. The feature vector for the body posture modality can be defined as follows:

$$FV_b = \{P_{1,x}(n), P_{1,y}(n), \ldots P_{14,y}(n), d(P_1(n), P_2(n)), \ldots, d(P_{13}(n), P_{14}(n))$$
$$, \theta\ P\ (n), P\ (n)\ , \ldots, \theta\ P\ (n), P\ (n) \quad (4)$$

The value $P_{i,c}(n)$ is the x, y co-ordinate value of i tracked point in the $n^{th}$ frame where c ϵ {x, y}. d is the Euclidean distance between two distinct tracked points. $\theta$ is the angle between two distinct tracked points.

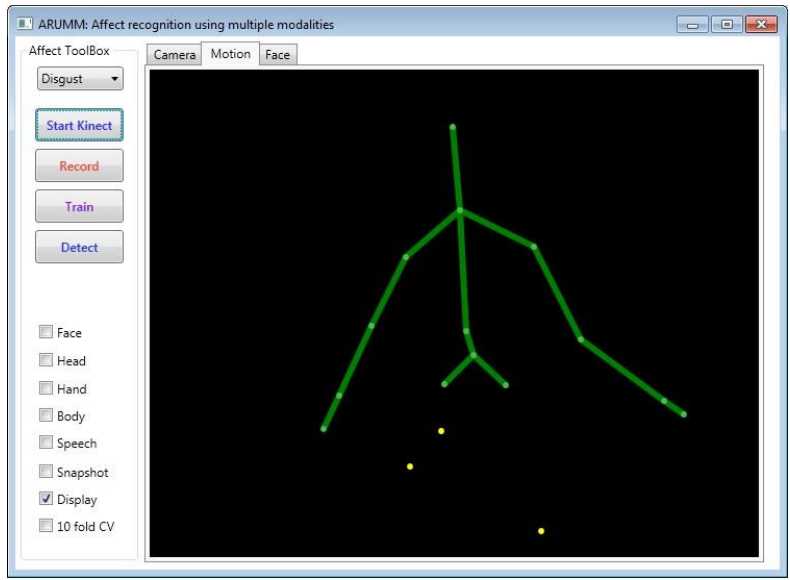

Figure 2: Skeletal joints tracking.

## 3. Image snapshot and speech analysis

### 3.1 Emotion Template Mapping Algorithm

The video input to the Kinect sensor is fed to the multimodal affect recognition system in the form of color, depth and skeletal data which is then processed by the tracking API for detecting facial points and the joints of hand and torso. In addition to this tracked points and joints, the research also introduces the concept of image snapshot. At any given point t, the input data from the skeleton contains frames of color images, skeletal data and depth data. The affect recognition system takes a snapshot of color frame at an interval of 5 seconds resulting in a png image. The interval was determined based on the processing time results discussed in detail in section 5. This image is then fed to the emotion template algorithm for recognizing affect from the face. For segmentation, we use skin color, adaptive threshold, on the face region detected using haar cascade classifier [18]. Mouth region is detected using Haar cascade classifier and serves as the region of interest for the template mapping algorithm. The edges on the mouth region are identified using canny edge detection. The library used for the segmentation is emgucv which is a c# wrapper for OpenCV. The mouth region is then analyzed by the affect recognition system using the emotion template algorithm. In this study, we developed an emotion template for detecting smile. The result of the image snapshot processing is then combined with the other modalities at the decision level as explained in the section 3.3. The rules used in the emotion template mapping algorithm to recognize smile from facial expression are described below:

$$e_l | |e_l| < t, e_{lx1} > m_{x1}, e_{ly1} > m_{y1} \tag{5}$$

$$e_r | |e_r| < t, e_{rx2} > m_{x2}, e_{ry2} > m_{y2} \tag{6}$$

$$e_n | \left(\frac{n_e}{N}\right) > n_t \tag{7}$$

$$t = \left(\frac{1}{8}\right) m_l, n_t = 0.05 \tag{8}$$

$e_{l,r}$ is an edge in the left and right half of the mouth region that has a length < threshold t. The values of t and $n_t$ that give best results using cross validation are shown in equation $e_{li1}, e_{ri2}, m_{i1}, m_{i2}$ are the location coordinates of the edge e and the bounded region of mouth m respectively and i ϵ {x, y}. The algorithm inspects edge count where $e_n$ is an edge that satisfies rule 1, rule 2 and rule 3 as per equation 5, 6, 7 respectively and classifies the mouth region as a region that contains a smile.

### 3.2 Key Word Lookup

The research proposes use of key word lookup technique for affect recognition from speech. The Microsoft speech recognition API is used for this purpose in combination with a set of key words associated with a certain class of emotion. The file is stored as an xml file containing rules and vocabulary for disgust, anger and happiness. The key word lookup based approach depends on the speech recognition accuracy. The speech API assigns a confidence level to each recognized word. The threshold used for the selecting a recognized word for further key word lookup is 0.3. The speech modality was currently only evaluated using English language. The final affect recognized based on the keyword lookup is then combined with outcome from other modalities at decision level. For this study, we evaluated disgust recognition using the keyword lookup technique. The set of words defined for recognizing disgust are shown below:

$$r_d \in \{nasty, foul, bad, ugly, hideous, awful, terrible, stink, pathetic, pitiful, sick, uggh, eeks, yuck\} \tag{9}$$

where $r_d$ is the recognized word with a confidence level > 0.3.

### 3.3 Decision Level Fusion

In this study a Support Vector Machine (SVM) based classifier was trained for each modality. Although the research focuses on 3 emotions (disgust, anger and happiness) out of the 6 basic emotions (anger, disgust, happiness, fear, sadness and surprise), the classifier for each modality was trained using data for all the 6 emotions. 5 different individuals enacted 6 different emotions in a single recording session for each emotion. Each session was 100 frames long. The multimodal affect recognition system was used to capture data for each modality resulting in a dataset per combination of emotion, modality and person. The input to the classifier was a feature vector using the tracked points, angles and Euclidean distance as discussed in the section 2 on a per frame basis. Thus, the output of the classifier for each modality can be generalized as a series of N predictions for an input session containing N frames where N = 100. The empirical probability of each emotion per N frames is given by:

$$P_e = \frac{n}{N} \mid e \, \varepsilon \, E \tag{10}$$

where n is the number of times emotion was identified as e by the classifier. The set E contains the six basic emotions denoted by a number 0 through 6. Anger was denoted as 0, happiness was denoted as 1, surprise was denoted as 2, disgust was denoted as 3, fear was denoted as 4, sadness was denoted as 5 and neutral was denoted as 6. Thus, the probability of each emotion measured for the test input frames was denoted by $P_0$, $P_1$, $P_2$, $P_3$, $P_4$, $P_5$, $P_6$ for anger, happiness, disgust, fear, sadness and neutral respectively. The final emotion identified for each modality is given by:

$$e = argmax \, P_e \tag{11}$$

The emotion per modality was detected for each frame and the resulting emotion from all the modalities was determined based on majority voting. For the emotion template mapping algorithm and the key word lookup from speech recognition the value of N was variable depending on the samples processed within the time it took for the other modalities to process 100 frames.

## 4. System prototype

The multimodal affect recognition system prototype was developed using c# as the scripting language and Windows Presentation Foundation (WPF). The image processing, classifier training and affect recognition was implemented using emgucv c# library. Facial expression and head tracking, hand and body position tracking was implemented using the skeletal tracking from Microsoft Kinect and Toolkit API. The user interface consists of viewing area for video, skeletal frame and facial expression and a toolbox area consisting of command buttons and checkboxes. The dropdown in the tool box can be used for selecting enacted emotions for training purpose. The voice input is available from the Kinect sensor and the keyword lookup based affect recognition is supported using Microsoft speech API.

The system functionality includes ability to record video and speech input, training classifier and detecting emotions real time. Each modality input, tracking and affect recognition can be switched on or off using the check boxes on the affect tool box. The system also has a checkbox to enable cross validation function that performs a 10-fold cross validation. The data is stored in text files containing comma separated feature values for each frame by emotion and modality used. Each line in the data file represents a feature vector. There is a button to start and stop the input from the Kinect sensor. The record button can be used to begin tracking the feature points in real time depending on the selected modalities. The detect button can be used for offline affect recognition from the captured data. The train button can be used to create SVM based classifier model. The three tabs Camera, Motion and Face are used as viewing areas for color video, hand and body positions tracking and face and head tracking respectively.

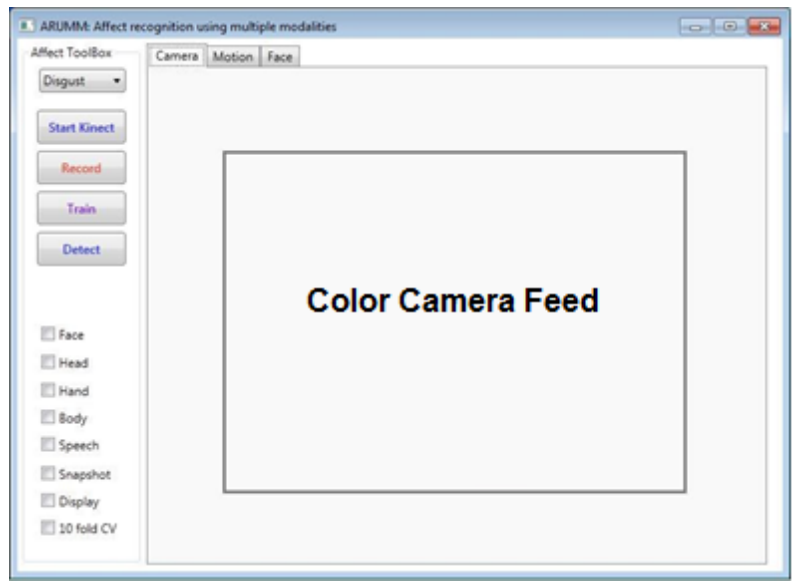

Figure 3: System prototype for affect recognition using multiple modalities.

## 5. Results

First the study discussed the offline affect recognition component results using classification rate, recall and precision. The evaluation was done using 10-fold cross validation. The videos were real time enacted video sessions in front of the Kinect sensor. All the enacted sessions were performed between 1.2 meter and 3.5 meters due to the range limitation on the sensor, in near frontal view and under controlled lighting conditions. The data from the Kinect sensor was stored as a comma separated text file containing the extracted feature vectors and the recognized class. It was ensured that each video recording fit the viewing screen area of the affect recognition system.

Table 1: Unimodal affect recognition results

| | | Classification Rate | Recall | Precision |
|---|---|---|---|---|
| Modalities | Facial Expression | 0.10, 0.26, 0.13 | 0.61, 1.57, 0.77 | 0.82, 0.73, 0.77 |
| | Head Position | 0.04, 0.18, 0.02 | 0.22, 1.07, 0.12 | 0.73, 0.86, 0.71 |
| | Hand Position | 0.05, 0.13, 0.08 | 0.3, 0.76, 0.48 | 0.66, 0.85, 0.81 |
| | Body Posture | 0.07, 0.11, 0.04 | 0.42, 0.67, 0.25 | 0.73, 0.78, 0.74 |
| | Template Mapping | 0.11 | 0.165 | 1 |
| | Speech | 0.065 | 0.43 | 0.70 |

The classification results are shown in Table 1 for individual modalities. Each column contains comma separated results for disgust, anger and smile respectively. Overall, the classification rate and recall rate is lower than expected. Also it was observed from the recall rate that the classifier had an affinity towards detecting anger as compared to other emotion classes. The precision rate values are modest and have scope for further improvement using multiple modalities. The template based mapping algorithm shows low recognition rate which indicates that for a more result oriented approach to yield good results compared to the traditional supervised learning techniques, it is important that the rules on which the algorithm depends are exhaustive. The CVL Face database [19] was used for validation of the template based mapping algorithm. The current algorithm would need to be improved before it can be used to add any value to the multimodal classification rates. The keyword lookup based approach for affect recognition from speech showed good precision results compared to other modalities. The lower recall rates could be attributed to the dependency on the speech recognition API and the accents of the test subjects.

Table 2: Multimodal affect recognition results

| | | Classification Rate | Recall | Precision |
|---|---|---|---|---|
| Emotions | Disgust | 0.12 | 0.42 | 0.92 |
| | Anger | 0.15 | 0.87 | 0.89 |
| | Smile | 0.09 | 0.54 | 0.87 |

Table 2 provides the classification results from the multimodal affect recognition system. The measurements agree with findings from existing research, that multiple modalities improve the recognition rates observed using unimodal affect recognition techniques. Moreover, the better precision rate in detecting disgust, anger and smile using multimodal affect recognition as compared to unimodal measurements indicate that the affect recognition system should be performed using multiple modalities instead of individual modalities for product feedback assessment.

Table 3: Affect recognition processing time in seconds/frame

| | | Seconds/frame |
|---|---|---|
| Modalities | Facial Expression | 0.03 |
| | Head Position | 0.021 |
| | Hand Position | 0.01 |
| | Body Posture | 0.014 |
| | Template Mapping | 3.8 |
| | Speech | NA |
| | Decision level fusion | 0.038 |

Secondly the study measures the processing time in seconds for affect detection using multiple modalities per frame. For each individual session, the decision level affect recognition processing time is within a second of the receiving outcomes from each modality as shown by the results in Table 3 which also provides results of affect recognition processing times for each individual modality and the total processing time. The total affect recognition processing time per frame can be calculated as follows:

$T_{total} = Max(t_{em}) + t_{fusion}$ (12) where $t_{em}$ is the time taken to recognize emotion e, using modality m and $t_{fusion}$ is the processing time taken for decision level fusion. The results indicate that only 26.31 tracked frames can be processed per second and this indicates that the current implementation of multimodal affect recognition component is better suited for offline execution. In the context of product feedback assessment, this is acceptable since the real time tracked features can be processed offline for feedback analysis.

Table 4: Feature tracking processing time in seconds

| | | Frames/sec | $T_{track}$/frame |
|---|---|---|---|
| Modalities | Facial Expression | 0.833 | 0.44 |
| | Head Position | 0.833 | 0.44 |
| | Hand Position | 1.7 | 0.225 |
| | Body Posture | 1.7 | 0.225 |
| | Template Mapping | 0.2 | NA |
| | Multiple modalities | 0.68 | 0.51 |

Finally the study measures the processing time (in seconds) for real time tracking of features using individual and multiple modalities. The metrics used was input frames received per second as shown in the first column and the average time $T_{track}$ taken for tracking per frame as shown in the second column in Table 4. In case of facial expression modality, the frequency of candidate frames fed as input to the system is < 1 frames/sec, and depends on underlying hardware and Kinect sensor processing time. For e.g. for a given input session for 20 seconds where a customer provides feedback, the actual number of candidate frames received and processed by system are < 20. For the keyword based lookup technique an average time of 0.25 milliseconds was measured per emotion recognition for performing the lookup. The multimodal affect recognition system prototype was implemented such that one component handles the processing of facial and head tracking and another component processes the hand and body posture tracking resulting in similar frames/sec and $T_{track}$/frame values for each pair of modality. The study measures the real-time performance by using the number of actual tracked frames for each modality and compares it with the number of actual tracked frames for multimodal system. The real-time tracking performance degrades but not significantly and it is still feasible to use the multimodal affect recognition system for real time tracking of data which can then be used for product feedback assessment in offline mode. The precision gain certainly outweighs the lesser processing time in multiple modalities since the affect recognition component can be used offline.

## 6. Conclusion

The study has shown that multimodal affect recognition is feasible in its application as product feedback assessment system, by performing disgust, anger and happiness recognition. The results also show that real time multimodal feature tracking and offline affect recognition is certainly feasible using Kinect sensor and has better accuracy compared to individual modalities [9]. The decision level fusion strategy provided high recognition results which agree with existing findings [17]. The low classification rates and recall rates indicate that more experiments need to be performed using additional features and more training data. The study did not find any evidence that using template based mapping algorithm and a keyword look up based approach for affect recognition from speech, can significantly improve the results of traditional supervised learning techniques at decision level. As a future scope, the effectiveness of the template mapping algorithm and keyword based lookup need to be investigated using more rules, vocabulary and other emotions in addition to smile and disgust. A prototype of multimodal affect recognition system was successfully implemented. The system provided functionality such as ability to capture features for facial expressions, voice, head and hand positions and body postures in real time. The system also provided ability to train classifiers and automatically detect affect using multiple modalities in offline mode.